# Magical Experience with Full-body Action

Bowen YE[1], Yuki ENZAKI [1], and Hiroo IWATA [1]

[1] *dept. of Data Science, Musashino University, Tokyo, Japan*

(Email: explosion0312@icloud.com)

**Abstract ---** This paper presents a system that generates a magical experience with full-body motion. The system consists of a locomotion interface and a spatial immersive display. A virtual experience system named the Magical Experience Generator was developed, equipped with a Magical Experience Controller. This system provides a physical movement experience along with magical-like interactions in a virtual space. We developed content inspired by the Japanese story "The Man Who Made Flowers Bloom" using Unity as the system's environment. The locomotion interface records the participant's walking trajectory and hand movements, representing their actions in the virtual space.

## 1 Introduction

We developed the Magical Experience Generator to create immersive VR experiences by combining physical movement with magical interactions. Many VR systems focus on visual and auditory stimuli but often overlook the integration of full-body motion, which significantly enhances the user's sense of presence and immersion. Full-body motion, especially walking, is a natural and intuitive method of locomotion in virtual environments, greatly enhancing immersion [1]. Current magical VR systems, such as "Waltz of the Wizard" and "The Unspoken" [2][3], primarily use hand gestures and controllers, limiting their ability to incorporate natural locomotion and full-body movement.

Several studies have explored the impact of full-body interactions in VR, with Liberatore and Wagner [4] highlighting how these interactions can significantly influence user attitudes and emphasizing the need for more immersive mechanisms. The Magical Experience Generator addresses these limitations by integrating walking and physical actions into the VR experience. Inspired by the Japanese story "The Man Who Made Flowers Bloom," users can interact with the virtual environment through physical movements and gestures, such as sprinkling virtual ashes to make withered trees bloom. The system uses a locomotion interface to simulate the sensation of walking and a base station to track hand movements, allowing users to perform magical actions that enhance immersion.

The primary objective of this research is to create a VR experience where walking and magical interactions are seamlessly integrated.

## 2 System Architecture of the Magical Experience Generator

The Magical Experience Generator is equipped with a spatial immersive display and a locomotion interface. The equipment used in this study includes an HTC VIVE Pro Eyes Full Kit (comprising one HMD, two controllers, and two base stations) and an Intel RealSense Depth Camera D455. The spatial immersive display provides 360-degree surrounding images not only horizontally but also vertically. The Torus Treadmill [1] is employed to provide a physical sense of walking. The RealSense is installed together with the Torus Treadmill for capturing walking motion. Additionally, a base station is used to track hand movements through a handheld controller, allowing for interactive experiences.

Information processing for the Magical Experience Generator is done by the Data Sensorium Server, which consists of the following modules: the Torus Treadmill Controller, the Digital Twin of Forest, and the Experience Database.

### 2.1 Torus Treadmill Controller

It receives motion data from the participant via the position sensor and calculates the position of the participant in the virtual forest environment. Activation of the floor of the Torus Treadmill is determined according to the displacement of the participant from the central position. The control signal is transmitted to the hardware of the Torus Treadmill. Additionally, the base station tracks hand movements through the handheld controller, enabling interactive actions such as throwing virtual objects. The motion data from the handheld controller is processed to provide a seamless and immersive interaction experience in the virtual environment.

### 2.2 Digital Twin of Forest

It is a 3D model of a virtual forest environment that includes data on the ground and trees. Scenes of the virtual forest are generated and transmitted to the spatial immersive display.

### 2.3 Experience Database

The position of the participant in the virtual forest environment is recorded and archived along with the time of the experience. Additionally, the database records the participant's movement trajectory, the trajectory of thrown objects, the position and time data from the handheld controller, interaction data, and the position and time data of thrown objects from generation to disappearance.

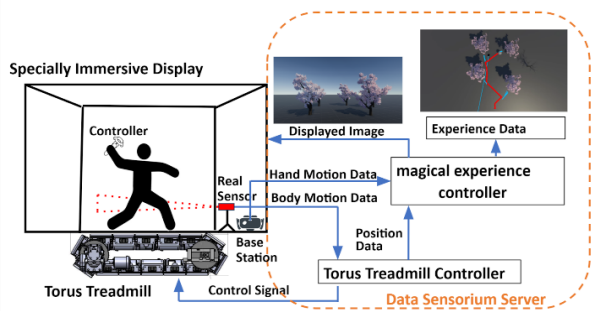

Fig.1　System Architecture of the Magical Experience Generator

## 3　SPATIAL IMMERSIVE DISPLAY

The Magical Experience Generator utilizes a spatial immersive display to present visual images of the virtual exhibition space. The display is a cuboid structure measuring 4 meters in width, 4 meters in depth, and 2.25 meters in height. Figure 2 shows the setup of the display. Participants enter the cuboid and physically walk within it. The images are projected onto the screens using eight projectors: four for the walls, two for the floor, and two for the ceiling. The wall screens are rear-projected, which prevents the participant's shadow from appearing on the screens. These projectors are capable of displaying stereoscopic images.

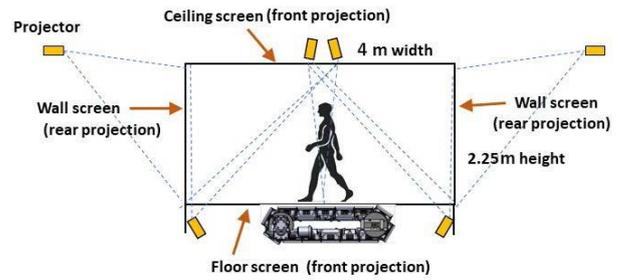

Fig.2　Projection system of the spatially immersive display

## 4　LOCOMOTION INTERFACE: TORUS TREADMILL

Torus Treadmill is a locomotion interface that creates a sense of walking. Although traveling on foot is the most intuitive way for locomotion, proprioceptive feedback from walking is not introduced in most applications of virtual environments. We have been developing an infinite surface driven by actuators for the creation of a sense of walking [5]. Torus-shaped surfaces are selected to realize the locomotion interface. The device employs 12 sets of treadmills. Figure 3 illustrates the basic structure of the Torus Treadmill. Each treadmill moves the walker along an "X" direction. These treadmills are connected side by side and driven in a perpendicular direction. This motion moves the walker along a "Y" direction. A combination of these motions enables the walker to omnidirectional walking. The walker can go in any direction while his/her position is fixed in the real world. Figure 4 shows the overall view of the device. The maximum walking speed of the device is 0.6m/s.

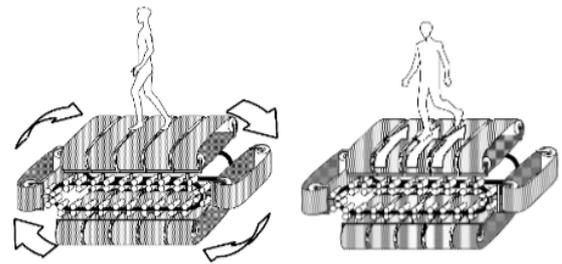

Fig.3　Structure of Torus Treadmill

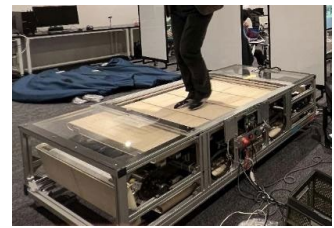

Fig.4　Overall view of Torus Treadmill

## 5　3D MODEL OF FANTAGIC WORLD

The story "The Man Who Made Flowers Bloom" is selected as the basis for creating an immersive and

interactive virtual experience. The 3D model for this magical experience is composed of three parts:

1. A series of tree models representing the transformation from withered trees to blooming trees, creating the forest scene.
2. A model of ashes, used to replicate the action of sprinkling ashes as described in the story.
3. Visualization models used to display the trajectories of the participant and the ashes in the log.

A script was developed for the tree models, enabling them to bloom gradually upon contact with the ashes model. Figure 5 shows the 3D model of the forest scene with various stages of tree transformation. Figure 6 shows the ashes model used for the interactive action of sprinkling ashes to make the trees bloom.

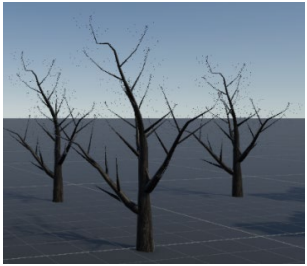

Fig.5　　Tree model

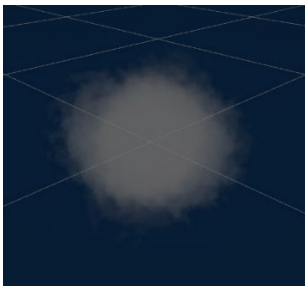

Fig.6　　Ashes model

Figure 7 shows a CG image of the 3D model. The forest scene is depicted after the ashes has been sprinkled on the trees, causing them to bloom as in the story.

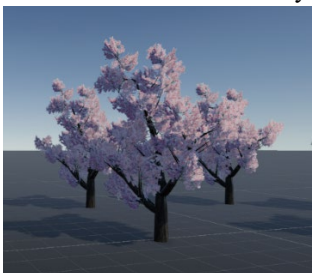

Fig.7　　CG Image of the 3D Model of the Magical Forest Scene

## 6 Results of Implementation and evaluation

A prototype of the Magical Experience Generator was implemented. Figure 8 shows the overall view of the prototype. Figure 9 shows the internal view of the setup.

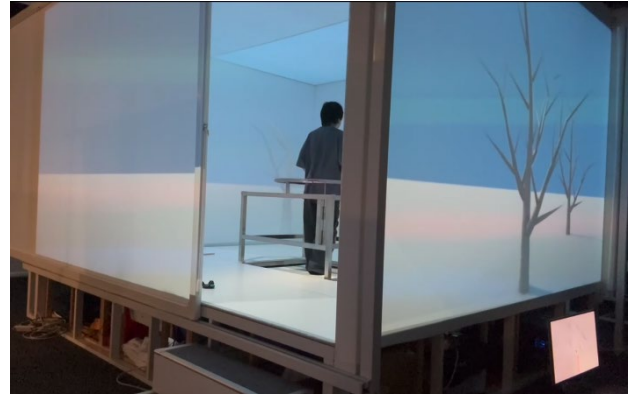

Fig.8　　Overall view of the prototype

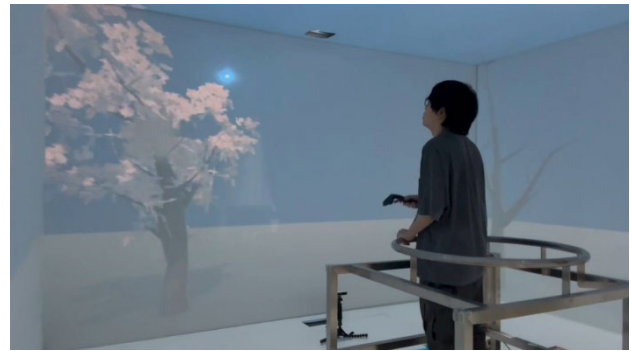

Fig.9　　Internal View of the prototype

To grab the ashes, participants needed to pull the trigger on the controller, and to throw the ashes, they had to release the trigger at the moment of throwing. The system detects the release of the trigger and calculates the throwing trajectory based on the motion data of the controller at that moment, allowing the ashes to follow a realistic parabolic path in the virtual environment.

The system recorded data every 0.1 seconds, resulting in 10 measurement frames per second. These frames captured the XYZ coordinates of both the participants and the handheld controller, as well as whether the trigger on the controller was pressed. This data was stored in CSV files for further analysis. The frequent data recording, indicated by the number of measurement frames, ensures a high level of detail in tracking the participants' movements and interactions within the virtual environment.

Figure 10 shows a 3D trajectory chart, displaying the movement paths of the controller and the target (the thrown ashes). Specifically, the figure illustrates the movement trajectories of the ashes and the controller, with blue lines representing the XYZ axis trajectories of the thrown ashes, and orange lines representing the XYZ axis trajectories of the handheld controller. From Figure 10, it is evident that the starting points of the ashes' movement paths align with the controller's trajectory, demonstrating that the ashes can be accurately thrown using the handheld controller.

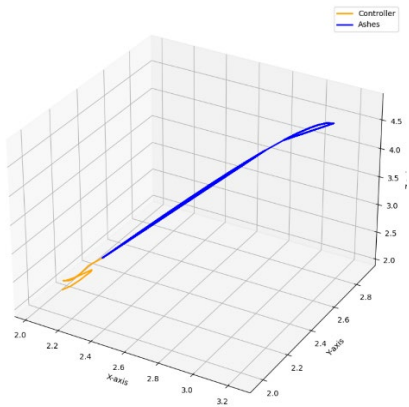

Fig.10   3D Trajectories of the Controller and Thrown Ashes

The data recorded in the CSV files allowed for the creation of visualizations that showed the participants' movement trajectories and the parabolic trajectories of the thrown ashes, providing clear evidence of the system's effectiveness in enabling natural interaction within the virtual environment. Figure 11 displays a visualization of the participants' movement trajectories and the ashes scattering activities. The red trajectories represent the movement paths of the participants, while the blue trajectories illustrate the paths of the scattered ashes. This evidence supports the effectiveness of the system in enabling natural interaction within the virtual environment.

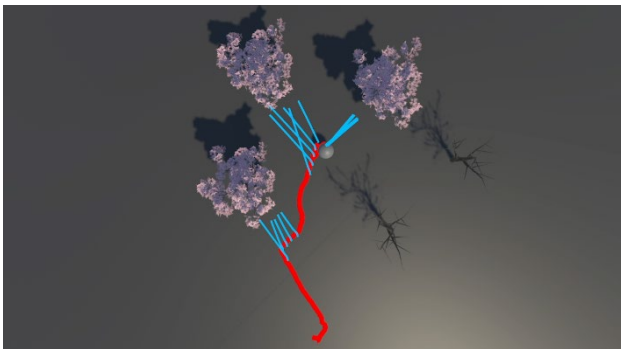

Fig.11   Visualization of Participants' Movement and Ashes Scattering Trajectories

## 7 Demonstration at Asia Haptics

The Magical Experience Generator will be demonstrated at the Asia Haptics conference. Due to the size and mobility constraints of the Torus Treadmill, it will not be used in the demonstration. Instead, the VIVE XR Elite kit (comprising one HMD and two controllers), along with the VIVE Ultimate Tracker 3+1 Kit, will be utilized. Together, these devices form the walk-in-place gesture input system, allowing attendees to perform walking motions in place, which are then translated into movement within the VR space. The demonstration will highlight the system's ability to provide an immersive and interactive magical experience by combining physical movement with virtual magical interactions. By clearly defining the components of the walk-in-place gesture input system, attendees will gain a comprehensive understanding of how the system replicates the sensation of walking in the virtual environment.

## 8 Conclusion

This paper has shown the basic concept and prototype implementation of the Magical Experience Generator, designed to recreate the magical experiences from the story "The Man Who Made Flowers Bloom." Its effectiveness in interactive engagement has been demonstrated through initial testing. Future work will include investigating the positive psychological impacts and potential benefits for mental health resulting from such immersive experiences. Additionally, efforts will continue to explore and develop more advanced and intuitive methods of interaction.

### Acknowledgement

This system is part of the Data Sensorium project, a conceptual framework designed to provide a physical experience of database content. It serves as a collaborative platform within Musashino University's Department of Data Science. A key feature is the integration of human action into the system architecture, enabling dynamic, interactive data experiences. For more details, refer to the related publication [5].